\documentclass[nologo]{pas}
\usepackage{multirow}
\usepackage{amsmath}
\usepackage{booktabs}
\usepackage{array}
\usepackage{siunitx}
\usepackage{tabularx}
\usepackage{ragged2e}

\begin{document}

\lefttitle{Sharp dynamic points in Earth-Sun physics}


\jnlPage{}{}
\jnlDoiYr{2025}
\doival{}

\articletitt{Research Paper}

\title{Sharp dynamic points in Earth-Sun physics}

\newif\ifdoubleblind
\doubleblindfalse  

\ifdoubleblind
 \author{
        \sn{Cambridge}, \gn{Author }$^{1,*}$, } 
 \affil{
        $^{1}$ A Pretty Nice Institution     } 

 \corresp{AUTHOR Cambridge, Email: author@cambridge.com}
        
\else
    \author{
        \sn{Rueda} \gn{José A.}$^{1,*}$,
        \sn{Ramírez} \gn{Sergio}$^{2}$,
        \sn{Sánchez} \gn{Miguel A.}$^{1}$,
        \sn{Aguilar} \gn{Cecilio U.}$^{1}$,
        and \sn{Rueda B.} \gn{Sandra}$^{3}$
    }

    \affil{
        $^{1}$Instituto de Agro-Ingeniería, Universidad del Papaloapan, 
        Loma Bonita, Oax. 68400, México; jrueda@unpa.edu.mx \\
        $^{2}$Facultad de Zootecnia y Ecología, Universidad Autónoma de Chihuahua, 
        Chih. 31000, México. \\
        $^{3}$Universidad Benito Juárez García, Real de Asientos, 
        Aguascalientes, 20710, México.
    }

    \corresp{José A. Rueda, Email: jrueda@unpa.edu.mx}

    \citeauth{
        José A. Rueda, Sergio Ramírez, Miguel A. Sánchez, Cecilio U. Aguilar, 
        and Sandra Rueda B. 
        Sharp dynamic points in Earth-Sun physics 
        {\it arXiv} 
        {\bf } 
    }
\fi

\history{(Received xx xx xxxx; revised xx xx xxxx; accepted xx xx xxxx)}

\begin{abstract}
The subsolar point, the closest location on Earth's surface to the Sun, marks the Sun-Earth line of gravity that governs Earth's coupled orbital-rotational cycle. We examined the dynamic interactions among the Sun meridian declination (SMD), the Equation of Time (EoT), Earth's rotational speed ($\mathrm{ER}_\omega$)---equatorial and with respect to the Sun--- and the path of the subsolar point (NBI) across longitude, including time derivatives up to the fourth order (snap). A central finding was that the function $\mathrm{NBI}_\alpha$(SMD) traces a lemniscate whose temporal structure mirrors the analemma, EoT(SMD), and whose symmetry converges to the obliquity component of the EoT. The EoT velocity ($\omega^*$) peaks at  solstices, troughs near the equinoxes, and crosses zero every mid-season.   $\mathrm{ER}_\omega$ decreases monotonically along trans-equinoctial phases where the net drives of EoT and SMD coincide, and increases along trans-solstitial phases, where their net drives oppose. Eight sharp kinematic periods were identified for the cycle SMD-EoT-$\mathrm{ER}_\omega$: two equinoctial, two solstitial, and one within each season. The non-solstitial sharp terms, defined by ZCPs and troughs of $\omega^*$, display a consistent 3$^\circ$ northward offset from the function $\mathrm{NBI}_\alpha$(SMD). These results reveal a direct dynamical link between SMD, EoT, and Earth's rotational speed, providing a novel framework for understanding Earth's rotation.
\end{abstract}


\maketitle
\section{Introduction}
Despite the fluctuations in Earth's rotational speed ($\mathrm{ER}_\omega$) being recognized as factual for timescales from daily to centennial, the prevailing theory sustains that Earth's rotation responds to the conservation of the angular momentum \citep{Gross2016}, overlooking the complexity of Sun-Earth physics. It is widely accepted that the Equation of Time (EoT) arises from the shape of Earth's orbit and the tilt of Earth's rotational axis \citep{Nurick2011, Raisz1941}. However, the role of the subsolar point in the dynamical interaction between the Sun meridian declination (SMD), the EoT, and $\mathrm{ER}_\omega$, remains unexplored. The current work assesses the combined dynamics and synchrony between SMD and EoT within a solar sundial noon analemma to establish a connection with $\mathrm{ER}_\omega$ along the Gregorian year.

When SMD is plotted against EoT, it produces the solar noon analemma, a lemniscate that describes the combined horizontal-vertical path of the mean-time noon Sun along the Gregorian year, as perceived from a specific location on the Earth’s surface \citep{Shaw2002}. Following the SMD within a solar noon analemma, the mean-time Sun travels from north to south and backward, depicting the y coordinate of the analemma or the Sun’s vertical path. Following the EoT within the solar noon analemma, the mean-time Sun travels from east to west and backward, depicting the x coordinate of the analemma, or the Sun’s horizontal path. 
The EoT establishes a rhythm for the daily deviation between the lengths of the mean day and the solar day \citep{Athayde2015}. Consequently, if the SMD is linked to $\mathrm{ER}_\omega$ along the planet's orbit, then the EoT reflects the fluctuation both in the length of the solar day (LSD) and in $\mathrm{ER}_\omega$. This hypothesis might seem bold, but the Sun's vertical path (SMD) and the Sun's horizontal path (EoT) are synchronized in such a flawless harmony that evidence supports causal in lieu of casual connection.

If natural beam irradiance (NBI) is expected to align with the local meridian at mean time noon, but the EoT occurs elsewhere, such a fact evidences a daily fluctuation in both the LSD and $\mathrm{ER}_\omega$. This work assumes that Sun-Earth physics defines $\mathrm{ER}_\omega$. Therefore, the angular momentum of rotation is considered as a consequence rather than the driver of Earth's rotation. Although LSD is tied to the $\mathrm{ER}_\omega$ \citep{Muller1995}, the light cone that supplies NBI to the subsolar point \citep{Rueda2024} comprises the clock hand that defines the only time frame Earth adheres to: solar time. Earth does not revolve around distant stars nor does it adhere to civil time, thus the actual length of the day should not be defined by distant stars, neither by a fixed 24-h timeframe. To comprehend the cause-effect relation between EoT and SMD, their interaction must be assessed in the context of the dynamics of the subsolar point at the true declination (SMD).

\section{Methods}
To analyze the joint vertical-horizontal path of the mean-time Sun, let $\delta$ denote the sun meridian declination (SMD) and $\delta^*$ the EoT (both in arcdeg). Thus, the joint function $\delta(\delta^*)$ displays the solar sundial noon analemma. The records of $\delta$ and $\delta^*$ correspond to the vertical and horizontal angles from the mean-time noon Sun to the Equator, or the local meridian, respectively. The EoT was expressed in spheric coordinates (arcdeg), to match the units of SMD. In this document, the EoT is referred as the Sun’s horizontal-path ($\delta^*$ ) while the SMD is referred as the Sun’s vertical-path ($\delta$).
The dynamical association between SMD and EoT was analyzed first, then $\mathrm{ER}_\omega$ was derived as the cause behind the EoT. Afterwards, the dynamics of the subsolar point (NBI dynamics) across longitude was assessed as the cause for the association between SMD and the EoT. The connection between SMD and EoT through the NBI path was proposed after observing that when the acceleration of NBI was plotted against SMD, it fits a lemniscate similar to the analemma.

\subsection{Parameters of the Sun's vertical path}
The SMD ($\delta$, equation~2) was derived from the fractional year (equation~1) modified for solar noon, following the analemma’s geometric model, coined by Fourier’s series \citep {Spencer1971}. Let $\delta$ denote the position of the Sun on its vertical path (SMD), which tracks the angle from the Equator to the Sun throughout the year. To explore the dynamics of the Sun’s vertical path in greater depth, five additional parameters were considered: angular velocity ($\omega$, $\mathrm{arcmin\,day^{-1}}$), angular acceleration ($\alpha$, $\mathrm{arcsec\,day^{-2}}$), angular jerk ($\xi$, $\mathrm{arcjerk\,day^{-3}}$), angular snap ($\varsigma$, $\mathrm{arcsnap\,day^{-4}}$), and net drive ($\zeta$) of SMD. The functions $\omega$, $\alpha$, $\xi$, and $\varsigma$ (equations 3 to 6) correspond to the first to fourth time derivatives of SMD ($\mathrm{d^k \delta/dt^k}$), in the same order. The net drive ($\zeta$, equation~7) was defined as the sign of the product $(\omega\alpha)$, where $\zeta$ is stated as accelerative (speeding up) when the sign is positive, or decelerative (slowing down) when the sign is negative.

\subsection{Parameters of the Sun’s horizontal path}
The EoT ($\delta^*$, equation~8) was derived from the fractional year (equation~1) modified for solar noon, following the geometric model \citep {Spencer1971} of the analemma. Let $\delta^*$ denote the position of the Sun on its horizontal path (EoT), which tracks the angle from the Sun to the local meridian throughout the Gregorian year. To explore the dynamics of the Sun’s horizontal path in greater depth, five additional parameters were analyzed: angular velocity ($\omega^*$, $\mathrm{arcmin\,day^{-1}}$), acceleration ($\alpha^*$, $\mathrm{arcsec\,day^{-2}}$), jerk ($\xi^*$, $\mathrm{arcjerk\,day^{-3}}$), angular snap ($\varsigma^*$, $\mathrm{arcsnap\,day^{-4}}$), and net drive ($\zeta^*$) of the EoT. The functions $\omega^*$, $\alpha^*$, $\xi^*$, and $\varsigma^*$ (equations 9 to 12) correspond to the first to fourth time derivatives of $\delta^*$, in the same order. The net drive of the EoT ($\zeta^*$, equation~13) is defined as the sign of the product $(\omega^*)(\alpha^*)$, where $\zeta^*$ is stated as accelerative (speeding up) when the sign of the given product is positive or decelerative (slowing down) when the sign is negative.

\begin{equation}
x = \frac{2\pi}{365}(t-1)
\end{equation}

\begin{multline}
\delta = \Big[0.006918 - 0.399912 \cos(x) + 0.070257 \sin(x) \\ - 0.006758 \cos(2x) 
+ 0.000907 \sin(2x) - 0.002697 \cos(3x) \\ + 0.00148 \sin(3x)\Big] \frac{180}{\pi}
\end{multline}

\begin{multline}
\omega = \frac{d\delta}{dt} = \frac{2\pi}{365} \Big[0.399912 \sin(x) + 0.070257 \cos(x) \\+ 0.013516 \sin(2x) + 0.001814 \cos(2x) + 0.00809 \sin(3x) \\+ 0.00444 \cos(3x)\Big] \frac{180}{\pi}
\left[\frac{\pi}{3}\frac{180}{\pi}\right]
\end{multline}

\begin{multline}
\alpha = \frac{d^2 \delta}{dt^2} = \left(\frac{2\pi}{365}\right)^2 \Big[0.399912 \cos(x) - 0.070257 \sin(x) \\+ 0.027032 \cos(2x)  - 0.003628 \sin(2x) + 0.024273 \cos(3x) \\ - 0.01332 \sin(3x)\Big] \frac{180}{\pi}
\left[\frac{\pi}{3}\frac{180}{\pi}\right]^2
\end{multline}

\begin{multline}
\xi = \frac{d^3 \delta}{dt^3} = \left(\frac{2\pi}{365}\right)^3 \Big[-0.399912 \sin(x) - 0.070257 \cos(x) \\- 0.054064 \sin(2x) - 0.007256 \cos(2x) - 0.072819 \sin(3x) \\ - 0.03996 \cos(3x)\Big] \frac{180}{\pi} 
\left[\frac{\pi}{3}\frac{180}{\pi}\right]^3
\end{multline}

\begin{multline}
\varsigma = \frac{d^4 \delta}{dt^4} = \left(\frac{2\pi}{365}\right)^4 \Big[-0.399912 \cos(x) + 0.070257 \sin(x) \\- 0.108128 \cos(2x) + 0.014512 \sin(2x) - 0.218457 \cos(3x) \\ + 0.11988 \sin(3x)\Big] \frac{180}{\pi}
\left[\frac{\pi}{3}\frac{180}{\pi}\right]^4
\end{multline}

\begin{equation}
\zeta = \mathrm{sign}(\omega \alpha)
\end{equation}

\begin{multline}
\delta^* = \mathrm{EoT} = -\Big[0.0000075 + 0.001868 \cos(x) \\ - 0.032077 \sin(x) - 0.014615 \cos(2x) \\ - 0.040849 \sin(2x)\Big] \frac{180}{\pi}
\end{multline}

\begin{multline}
\omega^* = \frac{d \delta^*}{dt} = -\frac{2\pi}{365} \Big[-0.001868 \sin(x) \\- 0.032077 \cos(x) 
+ 0.02923 \sin(2x) \\- 0.081698 \cos(2x)\Big] \frac{180}{\pi} 
\left[\frac{\pi}{3}\frac{180}{\pi}\right]
\end{multline}

\begin{multline}
\alpha^* = \frac{d^2 \delta^*}{dt^2} = -\left(\frac{2\pi}{365}\right)^2 \Big[-0.001868 \cos(x) \\+ 0.032077 \sin(x) 
+ 0.05846 \cos(2x) \\+ 0.163396 \sin(2x)\Big] \frac{180}{\pi} \left[\frac{\pi}{3}\frac{180}{\pi}\right]^2
\end{multline}

\begin{multline}
\xi^* = \frac{d^3 \delta^*}{dt^3} = -\left(\frac{2\pi}{365}\right)^3 \Big[0.001868 \sin(x) \\+ 0.032077 \cos(x) 
- 0.11692 \sin(2x) \\+ 0.326792 \cos(2x)\Big] \frac{180}{\pi} 
\left[\frac{\pi}{3}\frac{180}{\pi}\right]^3
\end{multline}

\begin{multline}
\varsigma^* = \frac{d^4 \delta^*}{dt^4} = -\left(\frac{2\pi}{365}\right)^4 \Big[0.001868 \cos(x) \\- 0.032077 \sin(x)
- 0.233840 \cos(2x) \\- 0.653584 \sin(2x)\Big] \frac{180}{\pi} 
\left[\frac{\pi}{3}\frac{180}{\pi}\right]^4
\end{multline}

\begin{equation}
\zeta^* = \mathrm{sign}(\omega^* \alpha^*)
\end{equation}

where $x$ is the fractional year in radians (equation~1), $t$ is time (day of the year, 1 to 365). 
The factors $[180/\pi]$ and $[(\pi/3)(180/\pi)]^2$ were applied to equations 3 to 6 and 9 to 12 
after the chain differentiation to switch the units from radians to arcdeg and to upscale the numerical 
records to agree with the proposed units, respectively. 
The negative sign in equations 8 to 12 switches the EoT from an on-sky analemma to a sundial noon analemma.

All signs of the horizontal path were arbitrarily reversed (equations 8 to 12)---from those arising from the geometric model of the EoT, to produce the solar sundial noon analemma. 
Inverting the signs of the EoT allowed intuitively associating negative records of $\delta^*$ to days where the Sun travels slower on its daily path across longitude, yielding an increased LSD (spring and autumn). 
Likewise, positive records of $\delta^*$ correspond to moments where the Sun travels faster on its daily path, yielding shorter LSD. Despite the ordinate axis of the sundial analemma (SMD) being also reversible, 
such inversion was unnecessary for the planned discussion.

\subsection{Adequate units for the time derivatives}
Because both the parameters of SMD and those of the EoT arise naturally in radians from the model, the factor $[180/\pi]$ was applied to equations 2 to 6 and equations 8 to 12, to express all derivatives in degrees of arc. When defining appropriate units for the parameters of the Sun's vertical path ($\delta$, $\omega$, $\alpha$, $\xi$, and $\varsigma$; equations 2 to 6), and later for those of the Sun's horizontal path ($\delta^*$, $\omega^*$, $\alpha^*$, $\xi^*$, and $\varsigma^*$; equations 8 to 12), the guiding principle was that every successive derivative reduced its range of variation up to approximately $1/(\pi/3) =  3/\pi$ radians, as compared to its integral. This ratio is equivalent to $1 / [(\pi/3)(180/\pi)]$ arcdeg, or just $1/60$ arcdeg. To account for the successive reduction in range of every next derivative, the $k$th time derivative of the EoT was scaled up by the factor $[(\pi/3)(180/\pi)]^k = 60^k$, aiming to avoid excessively small numerical values, while preserving the adequacy of the dimensionless scaling factor, when applied to quantities expressed in degrees of arc. $k$ indicates the order of differentiation. The two factors included in equations 2 to 6 and equation 8 to 12, must be interpreted cautiously. For instance, the derivatives were conducted before applying the factors, so the factors do not modify the differentiation chain, but its resultant derivatives, in order to adequate them for the units proposed later.

Because after applying the factor, the records of every successive derivative were scaled up by $60^k$ with respect its integral, it was only natural that the units of all the derivatives were scaled down, following the sexagesimal system---which is the best choice for Earth related coordinates. Once the factor $[180/\pi]$ was applied, all parameters were now given in degrees of arc. Therefore, after upscaling the numerical records, the most suitable units for position, velocity, acceleration, jerk, and snap were: arcdeg, arcmin day$^{-1}$, arcsec day$^{-2}$, arcjerk day$^{-3}$, and arcsnap day$^{-4}$, same order. Once the factors were applied and the units assigned, the records of all dynamical parameters of the SMD and EoT fell within the range $[-50 , 50]$. In order to establish the similarities between the dynamics of the SMD ($\delta$) and that of the EoT ($\delta^*$) analogous symbols and units were proposed for corresponding parameters.

Whether the positional parameters ($\delta$ and $\delta^*$) and their time derivatives were plotted against time, they would display pseudo sinusoidal functions \citep{Rueda2024}. Consequently, the terminology of sinusoidal curves is embraced from this point onward. In this context, the time interval in which a full cycle is completed is known as the period. A cycle can be divided in half-cycles, each framed between two zero-crossing points (ZCPs), two crests, or two troughs. Unlike the SMD and the EoT ($\delta$ and $\delta^*$), the crests and troughs of actual sinusoidal curves are equidistant to their ZCPs (abscissa), such distance referred to as amplitude. For practical purposes, the annual dynamics of the Sun's horizontal path must be divided in two cycles---each framed within a half-cycle of the Sun's vertical path (in either hemisphere), despite initial and final conditions of the EoT only converging by the end of the Gregorian year.

\subsection{Sections and phases of the analemma}
The Sun's vertical-path ($\delta$) was divided into two half-cycles by the ZCPs of $\delta$, which converge at the equinoxes with either a crest or a trough of $\omega$. Every half-cycle tracks the SMD within one hemisphere. Furthermore, dividing the SMD by the solstitial ZCPs of $\omega$ and equinoctial maxima of $\omega$, yielded quarter-cycles comprising an entire season.

To analyze the annual cycle of the Sun's horizontal path, the sundial analemma was divided in eight sections, whose boundaries are 8 key instances of EoT velocity ($\omega^*$): two troughs (near equinoxes), two crests (at solstices), and four midseason ZCPs of $\omega^*$. Let the eight sections be denoted early spring, late spring, early summer, late summer, early autumn, late autumn, early winter, and late winter. Let the section boundaries be spring equinox, midspring, summer solstice, midsummer, autumn equinox, midautumn, winter solstice and midwinter. The decision to divide the EoT by the 8 key instances of $\omega^*$ was based on the observation that these instances exhibit a consistently closer alignment with the equinoxes and solstices compared to the extrema and ZCPs of the EoT itself ($\delta^*$). For instance, $\omega^*$ troughs take place around +3 arcdeg of SMD, whereas $\delta^*$ crosses the meridian near +9 arcdeg of SMD; a six arcdeg closer alignment for the equinoctial boundaries.

To investigate the connection between the EoT and Earth's rotational speed, the analemma was also divided in two trans-equinoctial phases (denoted I and III) and two trans-solstitial phases (denoted II and IV), each phase including two consecutive sections of the analemma. The roman numbering obeys to the order in which the four phases occur. Unlike the seasons of the SMD, which consist of two sections with opposing horizontal directions (EoT direction), an analemmatic phase consists of two consecutive sections with consistent horizontal direction.

The direction of the EoT is identified by the sign of $\omega^*$. For instance, when $\omega^*$ is positive $\delta^*$ points right, but when $\omega^*$ negative $\delta^*$ points left. Each analemmatic phase extends between two ZCPs of $\omega^*$, while being homogeneous regarding: (1) the sign of $\omega^*$, (2) the direction of $\delta^*$, (3) the direction in $\mathrm{ER}_\omega$ dynamics and (4) LSD. The trans-equinoctial phase I includes late-winter and early-spring, and the trans-equinoctial phase III includes late summer and early autumn. The trans-solstitial phase II includes late spring and early summer, and the trans-solstitial phase IV encompasses late autumn and early winter.

\subsection{Earth’s rotational speed and the EoT}
Because the rotational speed of the Earth ($\mathrm{ER}_\omega$), is discussed alongside the vertical ($\omega$) and horizontal ($\omega^*$) velocities of the mean-time noon Sun, the concept of speed is reserved for the planet's rotation, which inherently implies $\mathrm{ER}_\omega$ is always a positive value.

As $\mathrm{ER}_\omega$ varies with latitude, it can be assessed by dividing Earth's circumference of a particular latitude by the 24 hours in a mean day. The length of the Equator is 40,075 km, a distance spanned by the apparent Sun every solar day. Accordingly, Earth's linear rotational speed averages 1669.78 km h$^{-1}$ at the Equator, or 27.8 km min$^{-1}$, where each degree of latitude spans 111.319 km. For the Tropics of Cancer and Capricorn, the same parameters correspond to 1532 km h$^{-1}$, 25.53 km min$^{-1}$ and 102 km, respectively.

An expression for $\mathrm{ER}_\omega$ was derived (equation~14) considering: (1) Earth's circumference (km) at the desired SMD, which signifies the distance the Sun spans across longitude within a day, given by $2\pi r \cos \delta$, (2) Earth's circumference shall be divided by 360 arcdeg, accounting for Earth's linear speed of rotation in km day$^{-1}$, (3) Earth's circumference must be adjusted by the EoT \citep{WilliamsND}, and finally (4) units of $\mathrm{ER}_\omega$ can be switched to km per hour dividing the outcome by 24. The radius of Earth's circumference used for the calculations was $r = 40075 \text{ km}/2\pi$, and the average speed of rotation was assessed as ER$\omega = (40075 \text{ km}/24 \text{ h})$.

Given that Earth's linear rotational speed varies with latitude equation~18 assesses Earth's equatorial speed of rotation (km h$^{-1}$) from the EoT. Equations 15 to 17 are the first to third time derivatives of $\mathrm{ER}_\omega$, which correspond to the acceleration, jerk and snap of Earth's rotation, respectively. Although the equations arise naturally in km per day, they were switched to km per hour by the factor given at the left of each formula, because $\mathrm{ER}_\omega$ is commonly stated in such units.

\begin{equation}
\mathrm{ER}\omega = \frac{1}{24} \left( \frac{2\pi r \cos \delta}{360} \right) (1 + \delta^*)
\end{equation}

\begin{equation}
\mathrm{ER}\alpha = \frac{d\rho}{dt} = \frac{1}{24^2} \left( \frac{2\pi r \cos \delta}{360} \right) \omega^*
\end{equation}

\begin{equation}
\mathrm{ER}\xi  = \frac{d^2 \rho}{dt^2} = \frac{1}{24^3} \left( \frac{2\pi r \cos \delta}{360} \right) \alpha^*
\end{equation}

\begin{equation}
\mathrm{ER}\varsigma = \frac{d^3 \rho}{dt^3} = \frac{1}{24^4} \left( \frac{2\pi r \cos \delta}{360} \right) \xi^*
\end{equation}

\begin{equation}
\rho_e = 1669.78 + 4.63827 \delta^*
\end{equation}

where $\mathrm{ER}_\omega$ is Earth's speed of rotation, while $\mathrm{ER}_\alpha$, ER$\xi$, and ER$\varsigma$ are the first to third time derivatives of $\mathrm{ER}_\omega$, and $\rho_e$ of equation~18 is the simplification of equation~14 for equatorial conditions. The units for equations 15 to 18 were switched from km day$^{-k}$ to km h$^{-1}$, km h$^{-2}$, km h$^{-3}$, and km h$^{-4}$, in that order, by means of the factor $(1 / 24^{(k+1)})$, which is not part of the differentiation chain but a multiplier required to handle a tricky parameter, which is given in km h$^{-1}$ but represents a day average. Finally, $\delta^*$ is the EoT in degrees of arc, while $\omega^*$, $\alpha^*$, and $\xi^*$ are the first to third time derivatives of $\delta^*$.

Earth's rotational speed depends entirely on the EoT, since it corresponds to a simple linear transformation of the $\delta^*$ with crests in midsummer and midwinter, and troughs in midspring and midautumn. To represent the equatorial speed of rotation: $\delta$ becomes zero and $\cos \delta$ yields 1, and $(2\pi r \cos \delta /360)$ gives 111.319 km arcdeg$^{-1}$, but the expressions were annotated in full in the equations 15 to 18 to explain their rationale. Along the present document, $\mathrm{ER}_\alpha$, ER$\xi$, and ER$\varsigma$ are the first to third time derivatives of $\mathrm{ER}_\omega$.

\subsection{The path of the subsolar point}
Because the shortest distance between two points is a straight line and because light travels in a straight line as well, the subsolar point is the nearest spot from the Sun compared to any other location on Earth's surface. Therefore, following the path of the subsolar point across longitude---which was denoted here as the path of natural beam irradiance (NBI)---might confirm the hypothesis that the solar declination, the Equation of Time and $\mathrm{ER}_\omega$ are inherently linked. In this context, $\mathrm{NBI}_\omega$ corresponds to the speed of NBI across longitude within a given solar day. The path of NBI was analyzed dismissing the EoT, to avoid forcing a relationship and to allow the natural association can be expressed. $\mathrm{NBI}_\omega$ (km day$^{-1}$) was derived as the distance in kilometers spanned by the subsolar point across longitude within a day, it can also be expressed in km h$^{-1}$, to match the units of $\mathrm{ER}_\omega$$\mathrm{ER}_\omega$.

Every day, NBI spans the entire planet across longitude on the latitude where the current SMD lands. Therefore, $\mathrm{NBI}_\omega$ was assessed as the length of Earth's circumference at a particular record of SMD (equation~19) which is the distance spanned by NBI within a solar day. To best describe the dynamics of the subsolar point, the equations 20 to 22 correspond to the first to third time derivatives of $\mathrm{NBI}_\omega$: acceleration, jerk and snap of NBI across longitude, respectively. 
\begin{equation}
\mathrm{NBI}_\omega = \frac{1}{24} \cdot 2\pi r \cos \delta
\end{equation}

\begin{equation}
\mathrm{NBI}_\alpha = \frac{d(\mathrm{NBI}_\omega)}{dt} = -\frac{1}{24^2} \cdot 2\pi r \omega \sin \delta
\end{equation}

\begin{equation}
\mathrm{NBI}_\xi = \frac{d^2 (\mathrm{NBI}_\omega)}{dt^2} = -\frac{1}{24^3} \cdot 2\pi r \left[ \omega^2 \cos \delta + \alpha \sin \delta \right]
\end{equation}

\begin{equation}
\begin{split}
\mathrm{NBI}_\varsigma = \frac{d^3 (\mathrm{NBI}_\omega)}{dt^3} = -\frac{1}{24^4} \cdot 2\pi r \big[ \xi \sin \delta \\ + 3\omega\alpha \cos \delta
- \omega^3 \sin \delta \big]
\end{split}
\end{equation}

where, $\mathrm{NBI}_\alpha$, $\mathrm{NBI}_\xi$, and $\mathrm{NBI}_\varsigma$ are the first to third time derivatives of the NBI speed ($\mathrm{NBI}_\omega$) in a given day, such day tagged by a particular solar meridian declination ($\delta$). $r$ is Earth's radius. $\delta$, $\omega$, $\xi$ and $\varsigma$ correspond to the sun meridian declination, as well as its velocity, acceleration and jerk. All SMD parameters in equations 19 to 22 shall be used here in radians, which required undoing the scaling up previously conducted by the factors.

By using SMD instead of latitude in equation~19, Earth's circumferences are evaluated only at latitudes where a SMD record exists, thereby tracing the path of the subsolar point (NBI) across longitude. Consequently, the distance between successive $\mathrm{NBI}_\omega$ records depends on $\omega$, the velocity of SMD. If the analysis of $\mathrm{NBI}_\omega$ reproduces the EoT---even partially---despite no EoT data being used in NBI assessment, this outcome would demonstrate a strong connection between SMD and the EoT, through NBI. Such a connection would support the central hypothesis of this study: Earth's rotational dynamics is inherently linked to the orbital dynamics. Data set was prepared and uploaded to a public repository \citep{Cambridge2025}.

\section{Results}
\subsection{The signs’ paradox}
The sign of the parameters describing the Sun's vertical path ($\delta$, $\omega$ and $\alpha$) arise from taking Northern Hemisphere as reference, which is customary. Nonetheless, all signs could be reversed without semantic consequence whether the Southern Hemisphere was the reference. In fact, the signs reverse naturally between hemispheres for all the dynamic parameters of SMD without interfering the seasonal net drive of SMD (speeding up or slowing down). Therefore, the direction in which the ZCPs of $\omega$ and $\alpha$ occur (downward or upward) is irrelevant from the standpoint of physics.

The parameters of the Sun's horizontal path ($\delta^*$, $\omega^*$, $\alpha^*$\ldots) cannot be reversed, because the meaning of a positive half-cycle of $\delta^*$ opposes that of a negative half-cycle. The inverse relationship between the length of the solar day and $\mathrm{ER}_\omega$ requires reversing the scale of the EoT for an intuitive analysis, as it has been considered for the fluctuations in the length of the planet's sideral day \citep{Georgieva2006}. The given inverse relationship motivated that the association between SMD and the EoT was based on the solar sundial noon analemma, where the direction of $\delta^*$ reverses from west to east and vice versa compared to the sky analemma. In this context, negative values of the EoT correspond to periods when $\mathrm{ER}_\omega$ falls below its annual average and the mean-time Sun lags behind its mean velocity. Conversely, positive values of the EoT correspond to periods when $\mathrm{ER}_\omega$ exceeds its annual average and the mean-time Sun runs ahead of its mean velocity. Following a sundial noon analemma, $\mathrm{ER}_\omega$ increases as the EoT advances to the right and decreases as the EoT progresses to the left. Within a solar sundial noon analemma, the mean-time Sun travels from left to right along the trans-solstitial phases and from right to left along the trans-equinoctial phases off the analemma.

Reversing the signs is advantageous when linking the EoT to both $\mathrm{ER}_\omega$ and the length of the solar day. For instance, at ZCPs of the EoT, the length of the solar day and $\mathrm{ER}_\omega$ converge around their annual averages (24 h and 1669.78 km h$^{-1}$). Nonetheless, a ZCP of $\delta^*$, can occur both within a trans-equinoctial phase---where $\mathrm{ER}_\omega$ decreases progressively---or within a trans-solstitial phase---where $\mathrm{ER}_\omega$ increases progressively, conforming a downward or upward ZCP, respectively.

When $\omega^*$ switches signs at a phase's boundary (midseason ZCPs of $\delta^*$), $\mathrm{ER}_\omega$ switches from a growing to a decreasing streak, or vice versa, while the dynamics of the length of the solar day reverses direction simultaneously, varying inversely with $\mathrm{ER}_\omega$.

\subsection{Dynamics of the Sun´s meridian declination }
The within season averages for position ($\delta$), angular velocity ($\omega$), acceleration ($\alpha$), jerk ($\xi$), snap ($\varsigma$) and net drive ($\zeta$) of the SMD are shown in Table~\ref{table:table1}. The period of the Sun's vertical-path ($\delta$) extends for one year, where its two half-cycles span the Northern and Southern Hemispheres, respectively. The crest and trough of $\omega$ occur near the equinoxes (at SMDs of 3.07 and 2.57$^\circ$), whereas the ZCPs of $\omega$ concur with the solstices.

The functions $\delta(\omega)$ and $\alpha(\omega)$ resemble circumferences. Accordingly, $|\omega|$ and $|\delta|$ vary inversely, so the highest $|\omega|$ occurs to the lowest $|\delta|$ and vice versa. An analogous association occurs between $|\omega|$ and $|\alpha|$. Therefore, $|\alpha|$ and $|\delta|$ vary directly.

After scaling up and assigning proper units, the seasonal averages of every parameter were similar between seasons. If signs were dismissed, season averages occur in the range of 12 to 20 for all the parameters of SMD. Weighted averages show a yearly equilibrium in SMD dynamics, as they approach zero, although a deviation of 0.356 arcdeg in SMD is indicative of some variable not considered. Data in (Table~\ref{table:table1}) must be interpreted with caution, as $\omega$, $\alpha$, $\xi$, and $\varsigma$ are stated in different units.

In boreal spring, positive records of $\delta$ and $\omega$ coincide with a negative $\alpha$, so the SMD decelerates as the apparent Sun departs from the Equator and approaches the Tropic of Cancer. In boreal summer, $\delta$ remains positive, while $\omega$ and $\alpha$ turn negative, so the SMD accelerates as the apparent Sun departs from the Tropic of Cancer and approaches the Equator. In boreal autumn, negative records of $\delta$ and $\omega$ coincide with a positive $\alpha$, so the SMD decelerates as the Sun travels from the Equator towards the Tropic of Capricorn. In boreal winter, $\delta$ remains negative while $\omega$ and $\alpha$ turn positive, so the SMD accelerates as the Sun moves from the Tropic of Capricorn toward the Equator.

For the Sun's vertical path, the net drive is decelerative when the signs of $\omega$ and $\alpha$ differ, spring or autumn, while accelerative when these signs coincide, summer or winter. As the signs of both $\omega$ and $\alpha$ are consistent within season, the same net drive endures along each season. Every particular $\delta$ holds characteristic records of $\omega$ and $\alpha$ along the Gregorian year, regardless of whether such association occurs within a decelerative or within an accelerative season, although differences do occur between hemispheres.

\subsection{Dynamics of the Equation of Time}
The position ($\delta^*$), angular velocity ($\omega^*$), acceleration ($\alpha^*$), jerk ($\xi^*$), snap ($\varsigma^*$) and net drive ($\zeta^*$) of the EoT, as well as the equatorial rotational speed of the Earth ($\rho$) are shown in (Table~\ref{table:table2}). The EoT reaches ZCPs---crosses through the local meridian---four times a year.

\begin{table*}
\caption{Parameters of the Sun's vertical path for every season of the year: solar declination ($\delta$), velocity ($\omega$), acceleration ($\alpha$), jerk ($\xi$), snap ($\varsigma$) and net drive ($\zeta$).}\label{table:table1}
{\tablefont\begin{tabular}{@{\extracolsep{15pt}}lccccccc}
\toprule
Season & Length & $\delta$ & $\omega$ & $\alpha$ & $\xi$ & $\varsigma$ & $\zeta$ \\
 & (days) & (arcdeg) & (arcmin day$^{-1}$) & (arcsec day$^{-2}$) & (arcjerk day$^{-3}$) & (arcsnap day$^{-4}$) & \\
\midrule
Spring & 93 & $+14.97$ & $+15.05$ & $-15.42$ & $-15.36$ & $14.21$ & $-$ \\
Summer & 93 & $+14.82$ & $-15.10$ & $-14.93$ & $15.42$ & $12.38$ & $+$ \\
Autumn & 90 & $-14.69$ & $-15.65$ & $+15.63$ & $19.42$ & $-13.02$ & $-$ \\
Winter & 89 & $-14.65$ & $+15.88$ & $+15.90$ & $-19.70$ & $-14.61$ & $+$ \\
\midrule
Weighted average & & $0.396$ & $0.000$ & $0.000$ & $0.000$ & $0.000$ & \\
\toprule
\end{tabular}}
\begin{tabnote}
{A negative net drive is decelerative, while a positive net drive is accelerative. The net drive is the sign of the product $\omega \alpha$.}
\end{tabnote}
\end{table*}   

\begin{table*}
\caption{Parameters of the Sun's horizontal path by analemmatic section: Equation of Time ($\delta^*$), velocity ($\omega^*$), acceleration ($\alpha^*$), jerk ($\xi^*$), snap ($\varsigma^*$), net drive ($\zeta^*$), and Earth's rotational speed ($\rho$).}\label{table:table2}
{\tablefont\begin{tabular}{@{\extracolsep{7pt}}lccccccccc}
\toprule
Period & Length & $\delta^*$ & $\omega^*$ & $\alpha^*$ & $\xi^*$ & $\varsigma^*$ & $\zeta^*$ & $\rho$ \\
 & (days) & (arcdeg) & (arcmin day$^{-1}$) & (arcsec day$^{-2}$) & (arcjerk day$^{-3}$) & (arcsnap day$^{-4}$) & & (km h$^{-1}$) \\
\midrule
Early spring & 46 & $-0.161$ & $-2.969$ & $6.212$ & $11.068$ & $-32.404$ & $-$ & $1669.2$ \\
Late spring & 39 & $-0.466$ & $2.148$ & $5.086$ & $-14.208$ & $-25.009$ & $+$ & $1667.5$ \\
Early summer & 35 & $1.195$ & $2.113$ & $-5.511$ & $-14.997$ & $23.957$ & $-$ & $1675.2$ \\
Late summer & 52 & $0.569$ & $-3.512$ & $-6.596$ & $10.972$ & $32.590$ & $+$ & $1672.4$ \\
Early autumn & 45 & $-3.108$ & $-3.598$ & $7.533$ & $15.849$ & $-26.036$ & $-$ & $1655.5$ \\
Late autumn & 51 & $-2.747$ & $4.385$ & $8.221$ & $-14.060$ & $-31.344$ & $+$ & $1657.3$ \\
Early winter & 53 & $2.139$ & $4.451$ & $-7.806$ & $-13.279$ & $31.900$ & $-$ & $1679.9$ \\
Late winter & 44 & $2.741$ & $-3.096$ & $-6.653$ & $16.066$ & $23.007$ & $+$ & $1682.6$ \\
\cmidrule(r){1-9}
Weighted average & 365 & $-0.0004$ & $0.000$ & $0.000$ & $0.000$ & $0.000$ & & $1669.8$ \\
\toprule
\end{tabular}}
\begin{tabnote}
{The last column is derived from $\delta^*$ in accordance with equation~18.}
\end{tabnote}
\end{table*}

For this reason, the annual dynamics of the EoT was divided into two cycles whose periods last 173 and 192 days, for the Northern and Southern Hemispheres, respectively.

Because the EoT was split by the key instances of $\omega^*$ rather than by those of $\delta^*$, the eight section boundaries---of the analemma--- consist of a trough, a crest, or a ZCP of $\omega^*$. Despite ZCPs of $\delta^*$ being the natural boundaries to divide the EoT, a most intuitive analysis derived by splitting $\delta^*$ from the key moments of $\omega^*$.

Dividing the EoT by the key instances of $\omega^*$ brought a number of advantages. Unlike the extrema of $\delta^*$, the crests of $\omega^*$ occur at the solstices and the troughs of $\omega^*$ near the equinoxes. Moreover, this approach allowed for the combined path $\delta(\delta^*)$ to be analyzed in sections whose horizontal net drive were consistent within but differing between. Thus, each cycle of the Sun's horizontal path includes the path of the mean time noon Sun within one hemisphere, although the equinoctial boundaries denoted spring equinox and autumn equinox occur at 3.07 and 2.57 arcdeg of SMD rather than at the Equator.

The dynamics of the Sun's horizontal path is analogous to that of the Sun's vertical path. Thus, (1) the maxima and ZCPs of $|\delta^*|$ and $|\alpha^*|$ occur nearby, therefore they vary directly, (2) $\delta^*$ and $\omega^*$ vary inversely within every section, (3) $\omega^*$ and $\alpha^*$ vary inversely, where the crests of $|\alpha^*|$ nearly coincide with ZCPs of $\omega^*$ and the crests of $|\omega^*|$ nearly meet the ZCPs of $\alpha^*$---so that $|\omega^*|$ and $|\alpha^*|$ cannot maximize together, and (4) initial and final conditions converge by the end of the Gregorian year, therefore the $\delta^*$ dual cycle repeats annually. Despite the synchrony between crests of $\omega^*$ and ZCPs of $\alpha^*$ or that between ZCPs of $\omega^*$ and crest/trough of $\delta^*$, $\alpha^*$ and $\delta^*$ do not maximize nor they reach ZCPs together.

The net drive of the EoT is accelerative for the four sections approaching the local meridian (labeled late-), but the net drive is always decelerative for the remaining four sections departing from the local meridian (labeled early), disregarding of the EoT (right or left).

The motion's direction is informed by the sign of $\omega^*$. A positive $\omega^*$ characterizes the trans-solstitial phases II and IV, where the EoT travels right. A negative $\omega^*$ characterizes the trans-equinoctial phases I and III, where the EoT travels left.

During solstices or equinoxes $\omega^*$ reaches an extrema in synchrony with a ZCPs of $\alpha^*$, but such coincidence holds an opposite meaning. At solstices, crests of $\omega^*$ (3.29 or 6.99 arcmin day$^{-1}$) converge to downward ZCPs of $\alpha^*$ five days after, or four days before, an upward ZCP of $\delta^*$, for the summer or winter solstice, respectively. On the other hand, the near equinoctial troughs of $\omega^*$ ($-7.786$ or $-5.65$ arcmin day$^{-1}$) meet an upward ZCPs of $\alpha^*$, 18 days before or 15 days after a downward ZCP of $\delta^*$.

\subsection{Combined horizontal-vertical path of the Sun}
Two cycles of the Sun's horizontal path (EoT) occur in synchrony with a unique cycle of the Sun's vertical path (SMD) along the Gregorian year. In the Northern Hemisphere, a cycle of $\delta^*$ meets a half-cycle of $\delta$, both comprising two seasons of the Gregorian year.

At midseason boundaries, $\omega^*$ reaches a ZCP in perfect synchrony with an extrema of $\delta^*$. In seasons whose net drive of SMD is accelerative, a downward ZCP of $\omega^*$ meets a crest of $\delta^*$, which takes place eight days before (midsummer) or six days after (midwinter) a trough of $\alpha^*$. In seasons whose net drive of SMD is decelerative, an upward ZCP of $\omega^*$ meets a trough of $\delta^*$, which takes place five days after (midspring) or four days before a crest of $\alpha^*$. The midseason boundaries, defined by ZCPs of $\omega^*$, converge to a similar analemmatic $\delta$ whether for the Northern ($-18.59$ or $-19.19$ for midspring and midsummer) or Southern Hemisphere ($-14.35$ or $-13.29$ for midautumn and midwinter), but such $\delta$ differs between them.

\begin{figure*}[p]
  \includegraphics[width=\textwidth, height=.9\textheight]{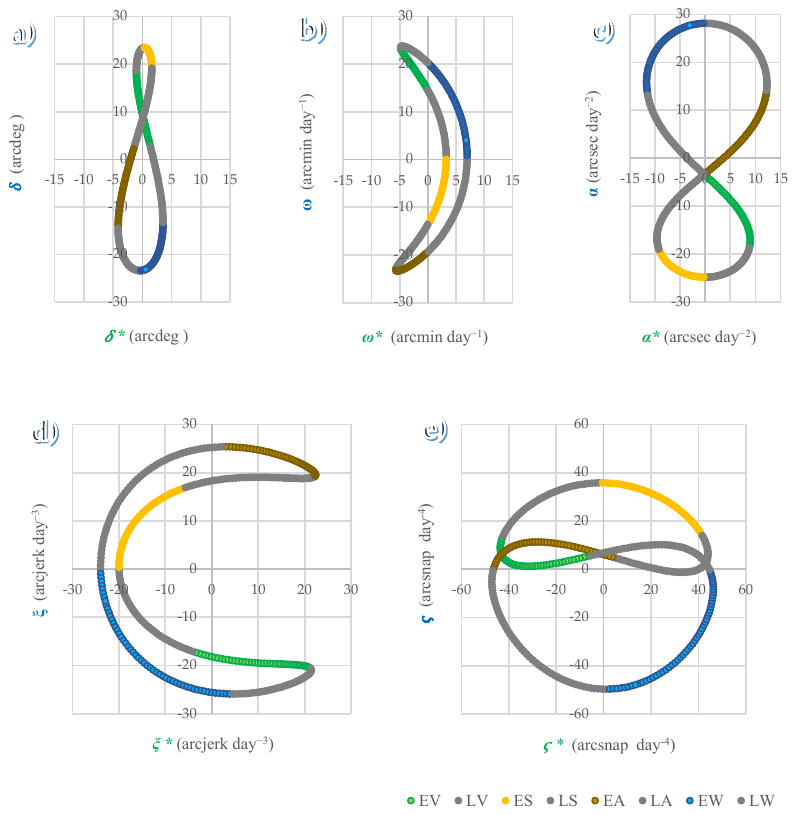}
  \caption{Association between the parameters of the Sun's vertical path (y-axis) and those of the Sun’s horizontal path (x-axis), including position (a), velocity (b), acceleration (c), jerk (d), and snap (e). EV:spring, LV: Late spring (V means vernal), ES: Early summer, LS: Late summer, EA: Early autumn, LA: Late autumn, EW: Early winter, and LW: Late winter. Every section denoted late- is colored grey to avoid overmarking.}
  \label{fig:fig1}
\end{figure*}

The direction of the mean time noon Sun regarding whether the EoT or the SMD is identified by the sign of the corresponding velocity. A positive $\omega$ indicates the mean-time Sun is moving toward the Tropic of Cancer (boreal direction), whereas a positive $\omega^*$ indicates the mean-time Sun is heading right. The association between the SMD and the EoT is summarized in Figure~\ref{fig:fig1}, where each parameter of Sun's vertical path is plotted against its corresponding parameter of the Sun's horizontal path. The ranges in which the velocity and acceleration of the SMD occur are twice larger than their corresponding parameters within the EoT, whereas jerk and snap vary within the same range for both the EoT and SMD. The parameters presented in Figure~\ref{fig:fig1} suggest a cause-effect association between SMD and EoT.

The net drives' analysis is presented in (Table~\ref{table:table3}). Regarding the net drive of the SMD, a season is accelerative when the mean time noon Sun approaches the Equator or decelerative when the mean time noon departs from the Equator. The net drive switches to decelerative when the SMD goes through a ZCP, such ZCP being the Equator. Every analemmatic section that departs de local meridian as well as every season (SMD) that departs from the Equator are decelerative. The signs of the parameters of the EoT occur in the very same order in either hemisphere. Although every parameter of the SMD reverses signs between hemispheres, the net drives of the SMD occur in the exact same order for the sequence of analemmatic sections of each hemisphere. Therefore, the same outcome would have arisen whether a positive declination was assigned to the Southern Hemisphere, as there is no up or down in the universe.

Every half-cycle of the EoT consists of two sections whose resultant drives oppose. According to the reversed $\delta^*$, each half-cycle of $\delta^*$ of the solar sundial noon analemma has a decelerative section followed by an accelerative section, such contrasting drives switching at a midseason trough of $\delta^*$. The second section of a season reverses the deviation from the local meridian caused by the first section; therefore, the Sun occurs near the local meridian around the end of the second section. The sign of the main three parameters (position, velocity and acceleration), whether for the EoT or SMD, remains consistent within every section.

The association between the EoT and SMD becomes clear even when the positional parameter of the apparent Sun is considered alone within the sundial noon analemma. For instance, the Pearson correlations between $\delta^*$ and $\delta$ yield 0.984 ($P<0.001$) for the trans-equinoctial phases and 0.905 ($P<0.001$) for the trans-solstitial phases. The association between the Sun's horizontal and Sun's vertical paths is also noticed in the coordination of their resultant drives in phases I and III, or by their opposition in phases II and IV.

Because the association fluctuates with SMD the analemma was sliced in four segments of SMD in order to compare the strength of the correlation between $\omega^*$ and $|\omega|$ for sections occurring in the same interval of SMD, but opposing directions of declination. Because of the signs' paradox, the sign of $\omega$ was dismissed. The Pearson correlation coefficients indicate that $\omega^*$ and $|\omega|$ correlate inversely throughout the Gregorian year.

The correlation coefficients between $\omega^*$ and $|\omega|$ were virtually equal for equivalent ranges of SMD (Figure~\ref{fig:fig1}b), including early spring vs late summer ($-0.997$, $-0.998$) as well as for early autumn vs late winter, ($-0.987$, $-0.987$). The high correlation remains when comparing late spring vs early summer ($-0.972$, $-0.962$) and late autumn vs early winter ($-0.963$, $-0.980$). Moreover, a crest of $\omega^*$ meets a ZCP of $\omega$ at each solstice; whereas a trough of $\omega^*$ takes place close to each equinoctial crests of $\omega$.

\begin{table*}
\caption{Signs of the main parameters and resultant drive of the Sun's horizontal and vertical paths (EoT and solar meridian declination) according to a solar sundial noon analemma, including: position ($\delta$ or $\delta^*$), velocity ($\omega$ or $\omega^*$), acceleration ($\alpha$ or $\alpha^*$) and net drive ($\zeta$ or $\zeta^*$).}\label{table:table3}
{\tablefont\begin{tabular}{@{\extracolsep{28pt}}lccccccccc}
\toprule
Section & \multicolumn{4}{c}{Sun's horizontal path} & \multicolumn{4}{c}{Sun's vertical path} & Combined \\
 & $\delta^*$ & $\omega^*$ & $\alpha^*$ & $\zeta^*$ & $\delta$ & $\omega$ & $\alpha$ & $\zeta$ & drive \\
\midrule
Early spring & $-$ & $-$ & $+$ & $-$ & $+$ & $+$ & $-$ & $-$ & Coordinated \\
Late spring & $-$ & $+$ & $+$ & $+$ & $+$ & $+$ & $-$ & $-$ & Opposed \\
Early summer & $+$ & $+$ & $-$ & $-$ & $+$ & $-$ & $-$ & $+$ & Opposed \\
Late summer & $+$ & $-$ & $-$ & $+$ & $+$ & $-$ & $-$ & $+$ & Coordinated \\
Early autumn & $-$ & $-$ & $+$ & $-$ & $-$ & $-$ & $+$ & $-$ & Coordinated \\
Late autumn & $-$ & $+$ & $+$ & $+$ & $-$ & $-$ & $+$ & $-$ & Opposed \\
Early winter & $+$ & $+$ & $-$ & $-$ & $-$ & $+$ & $+$ & $+$ & Opposed \\
Late winter & $+$ & $-$ & $-$ & $+$ & $-$ & $+$ & $+$ & $+$ & Coordinated \\
\bottomrule
\end{tabular}}
\begin{tabnote}
{A negative net drive is named decelerative and a positive net drive is named accelerative. The net drives of the EoT and SMD are coordinated when they align (trans-equinoctial phases) or opposed when they differ (trans-solstitial phases).}
\end{tabnote}
\end{table*}

The sections within a trans-solstitial phase display contrasting dynamics: the first section is characterized by an increasing $\omega^*$ and a decreasing $|\omega|$, while the second section is characterized by a decreasing $\omega^*$ and an increasing $|\omega|$. In the first section of each trans-equinoctial phase, $\omega^*$ and $|\omega|$ increase together as the combined direction of the mean-time Sun approaches simultaneously the Equator and the local meridian but decrease together along the second section of the phase while the direction of the mean-time Sun departs at once from the Equator and the local meridian. Despite the consistent increasing or decreasing pattern of $\alpha$ within each season (Figure~\ref{fig:fig1}c), $\alpha^*$ increases monotonically along the first section and decreases monotonically along the second section of every season, whereas the signs of both accelerations remain unchanged along the season. The horizontal and vertical accelerations, $\alpha^*$ and $\alpha$, vary directly along the first section of every season but inversely along the second. The Pearson correlation coefficients between $\alpha^*$ and $\alpha$ are virtually identical between the two sections of the same season, but their signs differ. This fact remains true when comparing early spring to late spring ($r = 0.96$ or $-0.93$), early summer to late summer ($r = 0.94$ or $-0.94$), early autumn to late autumn ($r = 0.90$ and $-0.90$), or early winter to late winter ($r = 0.89$ or $-0.88$). As the sign of $\alpha$ was dismissed for this analysis, the sign of the correlation coefficient obeys to changes in $\alpha^*$ between the two sections of the same season, because $\alpha^*$ switches direction at midseason.

A ZCP of $\alpha^*$ converges to the crests of $\alpha$ at the solstices, while the extrema of $\alpha^*$ (a crest or a trough) occur close to every midseason boundary at characteristic records of $\delta$, whereas a ZCP of $\alpha^*$ occurs near the vernal and autumnal equinoxes, at $\delta$ records of 2.928 and 2.714, respectively. Whether for the Sun's vertical or horizontal path, an extrema of acceleration converges with a change in direction, a behavior characteristic of pendular motion.

Given that the local meridian and the Equator conform the equilibrium points for the EoT and SMD, respectively, every shift departing from the equilibrium point is decelerative and every shift approaching the equilibrium point is accelerative, whether for the Sun's horizontal or vertical path. The net drive of a section becomes accelerative when the velocity is monotonically increasing or decelerative when the velocity is monotonically decreasing. An accelerative net drive can occur on either a positive or a negative direction, as long as the motion direction occurs towards the equilibrium point. Analogously, a decelerative net drive can occur through either a positive or a negative direction, as long as such direction departs from the equilibrium point. The direction of the actual motion can be retrieved from the sign of the velocity.

\subsection{Earth’s speed of rotation}
This document defines solar day as the unique true day and assumes the Sun's horizontal path is inherently tied to $\mathrm{ER}_\omega$. On a hypothetical day when the EoT runs 16 min ahead of the meridian, this offset corresponds to a horizontal angle of $+4$ arcdeg beyond the local meridian, equivalent to 240 arcmin per day, or 10 arcmin per hour compared to the mean day. This deviation indicates the Sun's horizontal path extends 445 km above the mean day and the equatorial rotational speed is 18.5 km hour$^{-1}$ faster than the annual average speed. The EoT ($\delta^*$), must be either added to or subtracted from the average speed of rotation (1669.78 km h$^{-1}$). In fact, $\mathrm{ER}_\omega$ varies from 1653 to 1688 km h$^{-1}$ at the Equator (or 27.5 to 28.0 km min$^{-1}$). A fluctuation involving $\pm 1\%$ throughout the year, equivalent to 852 km day$^{-1}$, 35.5 km h$^{-1}$, or 0.6 km min$^{-1}$ between the maximum and minimum speeds.

Dividing the EoT in sections allowed for a close examination of the within season and interseason net drives. Nonetheless, to analyze the dynamics behind the length of the solar day and Earth's rotational speed at the equator ($\mathrm{ER}_\omega$, or just $\rho$), a more effective analysis arises by dividing the analemma in four phases according to their horizontal direction, where each phase encompasses two successive sections of the EoT. The key moments of $\mathrm{ER}_\omega$ within a sundial noon analemma are inherited from the EoT. The crests and troughs of $\mathrm{ER}_\omega$ correspond to the crests and troughs of $\delta^*$---at ZCPs of $\omega^*$. For instance, the midspring and midautumn troughs of the EoT correspond to troughs of $\mathrm{ER}_\omega$ (1665.3 and 1650.8 km h$^{-1}$), whereas the midsummer and midwinter crests of the EoT correspond to crests of $\mathrm{ER}_\omega$ (1677.4 and 1686.3 km h$^{-1}$).

Earth's rotational speed accomplishes two phases of progressive decreases throughout the Gregorian year, denoted trans-equinoctial phases I and III. The phase I encompasses late winter and early spring, whereas the phase III encompasses late summer and early autumn. In the equinoctial phases I and III, $\mathrm{ER}_\omega$ behaves monotonically decreasing from a crest to a trough of $\delta^*$ (both including a ZCPs of $\omega^*$). In the trans-equinoctial phases, $\mathrm{ER}_\omega$ decreases despite the net drives of the $\delta^*$ sections being accelerative before the near equinoctial trough of $\omega^*$ and decelerative after the near equinoctial trough of $\omega^*$, yielding growing drops followed by decreasing drops in $\mathrm{ER}_\omega$, respectively.
The net drives of the EoT and SMD become opposed along the trans-solstitial phases II and IV. In the trans-solstitial phase II, an accelerative $\delta^*$ and a decelerative $\delta$ characterize late spring, but both net drives reverse for early summer. In the trans-solstitial phase IV an accelerative $\delta^*$ and a decelerative $\delta$ characterize late autumn, but the net drives of the EoT and SMD reverse for early winter. Earth-Sun dynamics causes $\mathrm{ER}_\omega$ to increase during the trans-solstitial phases of the EoT, for SMD records exceeding the SMD ranges parenthetically specified above.
The Earth rotates below its average speed during most of spring and autumn (seasons whose SMD is decelerative), but above its average speed during most of summer and winter (seasons whose SMD is accelerative). The average $\mathrm{ER}_\omega$ is reached only four times a year, either at the downward or upward ZCPs of the EoT ($\delta^*=0$), on 16 Apr, 15 Jun, 2 Sept, and 26 Dec (days 106, 166, 245 and 360 of the year); whereas the $\mathrm{ER}_\omega$ crests occur on 14 Feb and 28 Jul and the $\mathrm{ER}_\omega$ troughs fall on 15 May and 1 Nov. Accordingly, the trans-equinoctial phases I and II, where $\mathrm{ER}_\omega$ decreases monotonically, span from 14 Feb to 15 May and from 28 Jul to 1 Nov, respectively; each lasting three months. Hence, the trans-solstitial phases II and IV, where $\mathrm{ER}_\omega$ increases monotonically, span from 1 Nov to 14 Feb, and from 15 May to 28 July.

Earth's rotational speed accomplishes two phases of progressive increases throughout the Gregorian year, denoted trans-solstitial phases II and IV. The phase II encompasses late spring and early summer, and the phase IV encompasses late autumn and early winter. In the trans-solstitial phases (II and IV), $\mathrm{ER}_\omega$ behaves monotonically increasing from a trough to a crest of $\delta^*$ (both tagged by ZCPs of $\omega^*$). In the trans-solstitial phases, $\mathrm{ER}_\omega$ increases despite the net drive of the $\delta^*$ sections being accelerative before the solstice and decelerative after the solstice (going midway through an $\alpha^*$ ZCPs and a $\omega^*$ crest), showing growing increments followed by decreasing increments in $\mathrm{ER}_\omega$, respectively.

The net drives of EoT and SMD are coordinated throughout each trans-equinoctial phase. In the trans-equinoctial phase I, both exhibit a coordinated accelerative net drive in late winter ($\delta : -13.29, 3.08$, range 16.3) but a coordinated decelerative net drive in early spring ($\delta : 3.08$ to 18.59, range 15.5). In the trans-equinoctial phase II, both exhibit a coordinated accelerative net drive in late summer ($\delta : 19.19$ to  2.57, range 16.2) but a coordinated decelerative net drive in early autumn ($\delta : 2.57$ to $-14.35$, range 16.9).

As a simplified and practical conclusion, the EoT and the SMD exhibit coordinated net drives within the $\delta$ interval of $-13$ to 19 arcdeg, centered in $\delta = +3$. Consequently, each of the analemmatic trans-equinoctial phase spans approximately 16 arcdeg of SMD. According to the synchrony between the SMD and the EoT, the Sun influences significantly $\mathrm{ER}_\omega$. To begin with, Earth's rotational axis is a perpendicular projection to the Equator. Meanwhile the axis of the sunlight cone---extending from the Sun's center to the subsolar point---also conforms a normal projection to the Earth's rotational axis, a relationship that holds true throughout the year. The angular distance between the last vector and Earth's Equator is known as SMD. Because the SMD is faultlessly synchronized with the four seasons along Earth's orbit, the association here described between the $\mathrm{ER}_\omega$ and the SMD, may obey to the dynamic interaction between the SMD and Earth's revolution. For instance, the Sun-Earth gravity imposes a torque which periodically forces Earth's Equator into the ecliptic \citep{Torge2023}.

The increasing $\mathrm{ER}_\omega$ characteristic of trans-solstitial phases II and IV, at high SMD, suggests that the angle at which the Sun reaches Earth modifies the Sun's influence on $\mathrm{ER}_\omega$. This perspective implies NBI tags the axis of the Sun-Earth gravity, because it marks the shortest distance between the Sun's and Earth's surfaces, by landing at the subsolar point. The coordination in the net drives between SMD and EoT along the trans-equinoctial phases I and III suggests Earth resists rotation as NBI approaches the SMD $+3$ arcdeg. Conversely, the proximity of SMD to either the Tropic of Cancer or the Tropic of Capricorn enables a faster rotation. Thus, $\rho$ increases as the length of the parallel hosting the NBI is shorter, and decreases as such length grows.

Although the center of mass-density controlling SMD lies in the Equator, the latitude $+3$ arcdeg conforms the equilibrium center for the association between the SMD and $\mathrm{ER}_\omega$. Because the $\mathrm{ER}_\omega$ dynamics differs between hemispheres, it can be hypothesized that the higher share of continental land of the Northern Hemisphere modifies the effect of SMD over $\mathrm{ER}_\omega$. The midseason boundaries (midspring, midsummer, etc.) of the analemma, where $\omega^*=0$, define the beginning and ending of the four analemmatic phases on which the dynamics of the EoT and $\mathrm{ER}_\omega$ progress in a consistent direction.

\subsection{Periods of dynamic stress}

The association between SMD, NBI, EoT, and Earth's rotation can be elucidated by analyzing their combined dynamics within the boundaries of the EoT's sections, which are defined as periods of sharp dynamic stress. Sharp dynamic periods are brief intervals lasting one to ten days, during which a subset of the dynamic parameters of the EoT, Natural Beam Irradiance, and Earth's rotation, may exhibit a crest, a trough or a ZCP. Four such intervals are tied to the midseason section boundaries of the EoT, two occur around equinoxes and two more occur at the solstices.

\begin{figure*}[p]
\includegraphics[width=\textwidth, height=0.95\textheight]{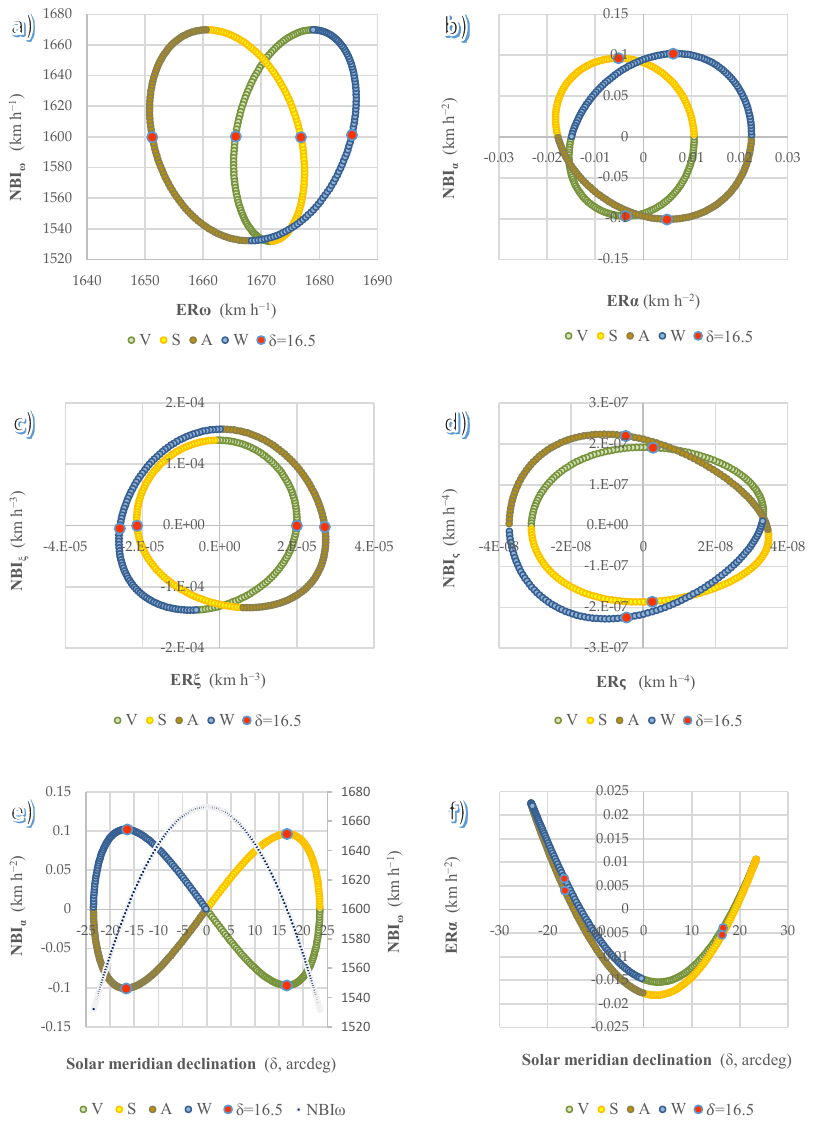}
  \caption{Dynamics of the subsolar point (here denoted as NBI) vs dynamics of Earth's rotation, involving speed (a),acceleration (b), jerk (c), and snap (d). Additionally, seasonal variation of $\mathrm{NBI}_\alpha$ (e), and $\mathrm{ER}_\alpha$ (f) are plotted against solar meridian declination.}
  \label{fig:fig2}
\end{figure*}

Table~\ref{table:table4} lists the key events of each sharp period, while Figure~\ref{fig:fig2} illustrates the relationship between subsolar point dynamics (NBI) and Earth's rotation ($\mathrm{ER}_\omega$) by comparing their speeds, accelerations, jerks, and snaps. 

The structure of Table~\ref{table:table4}   was designed to highlight the reverse timing patterns between pairs of sharp terms which share latitude. For upward SMD motion (midwinter, spring equinox, and midspring), events are presented in reverse chronological order (last goes first). For downward SMD motion (midsummer, autumn equinox, and midautumn), chronological order is maintained. The summer solstice sequence is also reversed due to its characteristic upward SMD bounce, while the winter solstice occurs within a downward pattern of SMD.

The sequence of events showed clear symmetries: spring and autumn equinoxes exhibit equivalent but reversed event sequences, as do summer and winter solstices. Midseason sharp terms mirror their hemispheric counterparts with which they share the  SMD sign---midspring mirrors midsummer, and midautumn mirrors midwinter---though with opposing EoT loops. For instance, midspring and midautumn (decelerative SMD) feature Earth's rotation troughs, while midsummer and midwinter (accelerative SMD) feature Earth's rotation crests.

These data consistently demonstrate the association between SMD, EoT, $\mathrm{NBI}_\alpha$, and $\mathrm{ER}_\omega$. While SMD governs the overall cycle (evident in equinoctial and solstitial sharp terms), the EoT strongly modulates midseason dynamics. The EoT, LSD, and $\mathrm{ER}_\omega$ thus represent complementary perspectives on a single unified cycle.

\subsubsection{Midseason sharp periods}

Midseason boundaries feature sharp dynamic intervals marked by extrema of $\delta^*$, $\alpha^*$, $\varsigma^*$, $\mathrm{ER}_\omega$, and $\mathrm{NBI}_\alpha$, as well as ZCPs in $\omega^*$ and $\mathrm{ER}_\alpha$. During midseason sharp periods, the association of EoT and $\rho$ with SMD is highlighted by the alignment of $\mathrm{ER}_\omega$ extrema with peaks of SMD jerk. For instance, midwinter hosts a $\mathrm{ER}_\omega$ crest preceded by an $\mathrm{NBI}_\alpha$ crest, which characterizes a season of accelerative SMD; conversely, midautumn hosts an $\mathrm{ER}_\omega$ trough followed by an $\mathrm{NBI}_\alpha$ trough, which characterizes a season of decelerative SMD.

When SMD is descending to the Tropic of Cancer (summer and autumn), the sharp period begins with an extremum of $\mathrm{ER}_\omega$ and ends with an extremum of $\mathrm{NBI}_\alpha$. Conversely, when SMD is ascending to the Tropic of Capricorn (winter and spring), the sharp interval begins with an extremum of $\mathrm{NBI}_\alpha$ and ends with an extremum of $\mathrm{ER}_\omega$.

\begin{table*}[htbp]
\caption{The eight sharpest periods in the combined dynamics Equation of Time / Earth’s speed of rotation / the path of the subsolar point (NBI).}
\label{table:table4}
\centering
\begin{tabular}{@{}r l l r l l @{} }
\toprule
\multicolumn{3}{c}{\textbf{Midspring}} & \multicolumn{3}{c}{\textbf{Midsummer}} \\
\midrule
135 & 15/5 & Trough $\mathrm{ER}_\omega$ & 209 & 28/7 & Crest $\mathrm{ER}_\omega$ \\
    &      & ZCPs: $\mathrm{ER}_\alpha, \omega^*$ &      &      & ZCPs: $\mathrm{ER}_\alpha, \omega^*$ \\
129 & 9/5 & Crests: $\alpha^*$, $\mathrm{ER}_\xi$, $\mathrm{NBI}_\varsigma$ & 217 & 5/8 & Troughs: $\alpha^*$, $\mathrm{ER}_\xi$, $\mathrm{NBI}_\varsigma$ \\
    &      & ZCPs: $\xi^*$, $\mathrm{ER}_\varsigma$ &      &      & ZCPs: $\xi^*$, $\mathrm{ER}_\varsigma$ \\
127 & 7/5 & Troughs: $\varsigma^*$, $\mathrm{NBI}_\alpha$ & 219 & 7/8 & Crests: $\varsigma^*$, $\mathrm{NBI}_\alpha$ \\
    &      & ZCP $\mathrm{NBI}_\xi$ &      &      & ZCP $\mathrm{NBI}_\xi$ \\
\\
\multicolumn{3}{c}{\textbf{Spring equinox}} & \multicolumn{3}{c}{\textbf{Autumn equinox}} \\
\midrule
88 & 29/3 & Troughs: $\mathrm{ER}_\alpha, \omega^*$ & 260 & 17/9 & Troughs: $\mathrm{ER}_\alpha, \omega^*$ \\
    &      & ZCPs: $\alpha^*$, $\mathrm{ER}_\xi$ &      &      & ZCPs: $\alpha^*$, $\mathrm{ER}_\xi$ \\
84 & 25/3 & Crest $\xi^*$, $\mathrm{ER}_\varsigma$ & 263 & 20/9 & Crests: $\xi^*$, $\mathrm{ER}_\varsigma$ \\
    &      & ZCP $\varsigma^*$ &      &      & ZCP $\varsigma^*$ \\
80 & 21/3 & Crest $\mathrm{NBI}_\omega$ ; ZCP $\mathrm{NBI}_\alpha$ & 267 & 24/9 & Crest $\mathrm{NBI}_\omega$ ; ZCP $\mathrm{NBI}_\alpha$ \\
78 & 19/3 & Trough $\mathrm{NBI}_\xi$ ; ZCP $\mathrm{NBI}_\varsigma$ & 269 & 26/9 & Trough $\mathrm{NBI}_\xi$ ; ZCP $\mathrm{NBI}_\varsigma$ \\
\\
\multicolumn{3}{c}{\textbf{Midwinter}} & \multicolumn{3}{c}{\textbf{Midautumn}} \\
\midrule
45 & 14/2 & Crest $\mathrm{ER}_\omega$ & 305 & 1/11 & Trough $\mathrm{ER}_\omega$ \\
    &      & ZCPs: $\mathrm{ER}_\alpha, \omega^*$ &      &      & ZCPs: $\mathrm{ER}_\alpha, \omega^*$ \\
39 & 8/2 & Troughs: $\alpha^*$, $\mathrm{ER}_\xi$, $\mathrm{NBI}_\varsigma$ & 309 & 5/11 & Crests: $\alpha^*$, $\mathrm{ER}_\xi$, $\mathrm{NBI}_\varsigma$ \\
    &      & ZCPs: $\xi^*$, $\mathrm{ER}_\varsigma$ &      &      & ZCPs: $\xi^*$, $\mathrm{ER}_\varsigma$ \\
37 & 6/2 & Crest $\varsigma^*$ & 310 & 6/11 & Trough $\varsigma^*$ \\
34 & 3/2 & Crest $\mathrm{NBI}_\alpha$ & 314 & 10/11 & Trough $\mathrm{NBI}_\alpha$ \\
    &      & ZCP $\mathrm{NBI}_\xi$ &      &      & ZCP $\mathrm{NBI}_\xi$ \\
\\
\multicolumn{3}{c}{\textbf{Summer solstice}} & \multicolumn{3}{c}{\textbf{Winter solstice}} \\
\midrule
173 & 22/6 & Crest $\mathrm{NBI}_\xi$ & 356 & 22/12 & Crest $\mathrm{NBI}_\xi$ \\
    &      & Troughs: $\xi^*$, $\mathrm{NBI}_\omega$, $\mathrm{ER}_\varsigma$ &      &      & Troughs: $\xi^*$, $\mathrm{NBI}_\omega$, $\mathrm{ER}_\varsigma$ \\
    &      & ZCPs: $\varsigma^*$, $\mathrm{NBI}_\alpha$ &      &      & ZCPs: $\varsigma^*$, $\mathrm{NBI}_\alpha$, $\mathrm{ER}_\xi$, $\mathrm{NBI}_\varsigma$ \\
172 & 21/6 & Crests: $\omega^*$, $\mathrm{ER}_\alpha$ & 357 & 23/12 & Crests: $\omega^*$, $\mathrm{ER}_\alpha$ \\
    &      & ZCPs: $\alpha^*$, $\mathrm{ER}_\xi$, $\mathrm{NBI}_\varsigma$ &      &      & ZCP $\alpha^*$ \\
\toprule
\end{tabular}
\begin{minipage}{\textwidth}
\justifying
$\omega^*$, $\alpha^*$, $\xi^*$ and $\varsigma^*$: velocity, acceleration, jerk and snap of the EoT ($\delta^*$). $\mathrm{NBI}_\omega$, $\mathrm{NBI}_\alpha$, $\mathrm{NBI}_\xi$ and $\mathrm{NBI}_\varsigma$: velocity, acceleration, jerk and snap of the subsolar point across longitude. $\mathrm{ER}_\omega$, $\mathrm{ER}_\alpha$, $\mathrm{ER}_\xi$ and $\mathrm{ER}_\varsigma$: velocity, acceleration, jerk and snap of Earth's equatorial rotation. DOY: Day of the year.
\end{minipage}
\end{table*}

\subsection{Kinematic analysis}

When SMD is departing from the Equator (midspring or midautumn), the sharp midseason period features troughs in $\varsigma^*$, $\mathrm{ER}_\omega$, and $\mathrm{NBI}_\alpha$, along with a crest in $\alpha^*$. In (southern SMD), the sharp midseason periods occur within similar SMD ranges: midautumn spans from $-14.18$ to $-16.94$ arcdeg, and midwinter from $-13.29$ to $-16.47$ arcdeg.

When SMD is approaching the Equator (in midsummer or midwinter), the sharp midseason period features crests in $\varsigma^*$, $\mathrm{ER}_\omega$, and $\mathrm{NBI}_\alpha$, along with a trough in $\alpha^*$. In spring and summer (northern SMD), the sharp midseason periods occur within similar SMD ranges: midspring spans from 16.86 to 18.67 arcdeg, and midsummer from 16.64 to 19.17 arcdeg. 

The sharp midseason periods of both boreal and austral seasons are framed between an extremum of $\rho$ and an extremum of $\mathrm{NBI}_\alpha$. The extrema of $\mathrm{NBI}_\alpha$ occur at the SMDs of: 16.86, 16.64, $-16.93$, and $-16.47$ arcdeg, on days 128, 219, 314 and 35 of the Gregorian year. These SMDs hold both the highest SMD jerk, as well as ZCPs of SMD snap.

Conversely, the crests and troughs in $\mathrm{ER}_\omega$ occur at 1.81, 2.53, 2.76, and 3.18 arcdeg northward from the SMDs where the referred $\mathrm{NBI}_\alpha$ extrema take place. The crest of $\mathrm{ER}_\omega$ occurs 10 days before the crest of $\mathrm{NBI}_\alpha$ in midsummer or 10 days after the crest of $\mathrm{NBI}_\alpha$ in midwinter, while the trough of $\mathrm{ER}_\omega$ occurs 7 days after the trough of $\mathrm{NBI}_\alpha$ in midspring or 9 days before the trough of $\mathrm{NBI}_\alpha$ in midautumn.

\subsubsection{Equinoctial and solstitial sharp periods}

Equinoxes trigger sharp periods involving crests of $\xi^*$ and $\mathrm{NBI}_\omega$, troughs in $\omega^*$ and $\mathrm{ER}_\alpha$, as well as ZCPs of $\alpha^*$, $\varsigma^*$, and $\mathrm{NBI}_\alpha$. The equinoctial sharp periods occur within SMD ranges of 0 to 3.07 arcdeg in the spring equinox, or $-0.14$ to 2.52 in the autumn equinox, the length of which being of 10 and 7 days, respectively.

Solstices trigger sharp dynamic intervals involving crests in $\omega^*$ and $\mathrm{ER}_\alpha$, troughs in $\xi^*$ and $\mathrm{NBI}_\omega$, as well as ZCPs of $\alpha^*$, $\varsigma^*$ and $\mathrm{NBI}_\alpha$. The solstitial sharp periods involve extrema for the same parameters as equinoctial sharp periods, except that crests and troughs are reversed.

The results of the present work oppose prevailing knowledge on Earth's rotational dynamics. For instance, a widely cited study summarizes $\mathrm{ER}_\omega$ dynamics across multiple timescales and proposes that seasonal variations in the length of the day account for only a few milliseconds \citep{Dickey1995}. However, that study was based on sidereal rather than solar days, and thus does not examine $\mathrm{ER}_\omega$ in the context of Sun-Earth physics. In our analysis, the sequence of dynamical events defining the sharp periods reverses between ascending and descending SMD. This inversion holds when comparing the sharp periods between boreal seasons, between austral seasons, and across solstices and equinoxes (Table~\ref{table:table3}).

\subsubsection{NBI acceleration as a driver of Earth's rotation}

Figure~\ref{fig:fig2} (panels a, b, c, and d) reveals a consistent kinematic link between NBI and $\mathrm{ER}_\omega$ across four levels of motion: velocity ($\omega$), acceleration ($\alpha$), jerk ($\xi$), and snap ($\varsigma$). A significant finding is that the midseason kinematic sharp periods converge within a narrow range. Their southward boundaries occur at either 16.5 or $-16.5$ arcdeg of SMD. The sharp periods are featured  by: (1) an extrema in $\mathrm{ER}_\omega$ converges with average records of $\mathrm{NBI}_\omega$, (2) extrema in  ($\mathrm{ER}_\xi$) aligns with ZCPs of $\mathrm{ER}_\xi$), (3) extrema of $\mathrm{NBI}_\alpha$ meets ZCPs of  $\mathrm{ER}_\alpha$, and (4) extrema of $\mathrm{NBI}_\varsigma$ falls near ZCPs of ($\mathrm{ER}_\varsigma$). This consistent, alternating lock-step is the definitive kinematic signature of a coupled system.

Because $\mathrm{NBI}_\omega$ is directly related to Earth's circumference at each latitude, it inherently peaks at equinoxes and troughs at solstices, while $\mathrm{NBI}_\xi$ exhibits the opposite pattern. $\mathrm{ER}_\alpha$ and $\mathrm{NBI}_\alpha$ exhibit an alternating direct/inverse relationship across successive sections of the sundial analemma, preventing simultaneous maxima. 
$\mathrm{NBI}_\alpha$ ranges from $-0.101$ to $0.102$ km h$^{-2}$ and $\mathrm{ER}_\alpha$ ranges from $-0.018$ to $0.022$ km h$^{-2}$. $\mathrm{ER}_\alpha$ peaks at solstices and troughs at equinoxes, while $\mathrm{NBI}_\alpha$ peaks and troughs in the midseason sharp periods (Figure~\ref{fig:fig2}b).This specific phase relationship between velocity and acceleration is a fundamental characteristic of the harmonic motion that defines all four cycles analyzed here: SMD, EoT, $\mathrm{ER}\omega$, and $\mathrm{NBI}\omega$.

Contrary to the expectation of a simple linear relationship---where NBI parameters would vary proportionally with the decreasing length of Earth's circumference from the Equator to the Tropics---the joint function $\mathrm{NBI}_\alpha$(SMD) traced a lemniscate that corresponds closely to the Equation of Time (EoT) in shape, timing, and direction (Figure~\ref{fig:fig2}e). The similarity in phase and structure between the $\mathrm{NBI}_\alpha$ and EoT lemniscates provided strong evidence for the SMD-EoT-$\mathrm{ER}_\omega$ connection. In fact, the function $\mathrm{NBI}_\alpha$($\mathrm\delta$) gives the causal explanation for the obliquity component of the EoT, such element described by \citet{Lynch2012}. The obliquity displays ascending ZCPs at equinoxes and descending ZCPs at the solstices \citep{Camuffo2021}, takes a vertically and horizontally symmetric shape centered at the Equator that resembles the function $\mathrm{NBI}_\alpha$($\mathrm\delta$). 
This finding rests on two methodological constraints: (1) $\mathrm{NBI}_\omega$ was derived within SMD  (true declination), and (2) the EoT was explicitly excluded when deriving  NBI, isolating the effect of SMD both on the EoT and $\mathrm{ER}_\omega$. 

Figure~\ref{fig:fig2}f further demonstrates that $\mathrm{ER}_\omega$ and SMD are linked through NBI, with $\mathrm{ER}_\omega$ displaying decelerative trends during trans-equinoctial phases and accelerative trends during trans-solstitial phases. A systematic discrepancy is nevertheless observed, in the form of a persistent northward offset of approximately 3 arcdeg. The equinoctial boundary of the analemmatic sections, defined by the minima of EoT velocity, lies 3 arcdeg north of the Equator, while midseason boundaries of the analemmatic sections also occur 3 arcdeg north of the expected 16.5 SMD.

This consistent displacement indicates that the influence of SMD on the EoT is systematically shifted northward, centered 3 arcdeg above the Equator, as clearly illustrated by the location of the equinoctial phases of the analemma. The geophysical reason for this persistent discrepancy remains unknown and poses a key question for future research.
This study establishes a clear framework linking Natural Beam Irradiance (NBI) to Earth’s rotation ($\mathrm{ER}_\omega$), addressing two fundamental questions: what drives Earth’s rotation, and where the Sun–Earth axis of gravity governing revolution is located.

This standpoint rests on two geometric premises: (1) two spheres can approach at only one point on their surfaces, and (2) since light travels along the shortest path between two points, the connection between the \textit{helioterric line of gravity} and \textit{Earth’s rotation} emerges through the path of the subsolar point, identified by Natural Beam Irradiance (NBI).

The subsolar point marks the shortest Sun–Earth distance, where their surfaces and centers of mass align. It is not merely a geometric abstraction but the physical locus where NBI is delivered, defining the pivot for both Earth’s revolution and rotation, with the Sun as sole driver. The analysis links orbital dynamics, irradiance geometry, and gravitational mechanics into a unique coherent theory.

This study concerns Earth’s rotation as defined by the length of the solar day, the interval between successive solar noons \citep{Meeus1998}. This differs from the conventional “length of day” (LOD), which measures rotation relative to atomic or sidereal references. In this framework, Earth’s rotational behaviour is interpreted entirely within the Sun–Earth geometric system, where Natural Beam Irradiance (NBI) defines the coupling axis and governs both rotation and revolution.

\subsection{Conclusion}

When viewed through the lens of the solar sundial noon analemma, the Sun meridian declination (SMD) emerges as the root explanatory parameter underlying both the Equation of Time (EoT) and Earth's rotational speed ($\mathrm{ER}_\omega$), wherein Earth's rotational dynamics can be understood as a reframing of the EoT that accounts for the dynamics in the length of the solar day (LSD). These four cycles are synchronized throughout Earth's orbit as a smooth dynamic, interconnected system. As the analemma's components follow pendular motion, the Equator acts as the equilibrium point for SMD, while the local meridian serves as the equilibrium point for the EoT. In both cases, accelerative motion occurs when the Sun approaches the equilibrium point, and decelerative motion when the Sun departs from the equilibrium point.

The trans-equinoctial phases I and III ($-13$ to 19 arcdeg of SMD, centered at $+3$), span about 32 arcdeg each and correspond to periods when the analemma progresses from right to left, a direction marked by negative EoT velocity. During these phases, the net drives of the SMD and EoT are coordinated, leading to reductions in $\mathrm{ER}_\omega$ and increases in LSD. Conversely, trans-solstitial phases II and IV (occurring from 19 to 23.5 and $-23.5$ to $-13$ arcdeg of SMD, spanning 4.5 or 10.5 arcdeg, same order) are periods when the analemma progresses from left to right, a direction marked by positive EoT velocity. During these phases the net drives oppose one another, leading to increases in $\mathrm{ER}_\omega$, and corresponding reductions in LSD.

When the acceleration of the subsolar point across longitude is plotted against SMD, such function denoted $\mathrm{NBI}_\alpha$(SMD), it traces a lemniscate whose temporal phasing closely mirrors that of the analemma. Given the fundamental role of acceleration in motion, the likeness between the analemma and the $\mathrm{NBI}_\alpha$(SMD) lemniscates serves as definitive proof of a cause-effect relationship between SMD and EoT. This, in turn, establishes a direct causal link between SMD and $\mathrm{ER}_\omega$; which means a solid frame to comprehend why does Earth rotate.

\section*{Concepts}
\textit{Amplitude}. In sinusoidal curves, the amplitude is the vertical distance from either the crest or the trough to the horizontal axis.

\textit{Angular velocity of SMD} ($\omega=d\delta/dt$, arcmin day$^{-1}$). The first-order time derivative of SMD, within a solar meridional analemma, which describes the rate of change for SMD.

\textit{Angular acceleration of SMD} ($\alpha=d\omega/dt$, arcsec day$^{-2}$). The second-order time derivative of SMD within a solar meridional analemma, which quantifies the rate of change for the velocity of SMD.

\textit{Angular jerk of SMD} ($\xi=d\alpha/dt$, arcjerk day$^{-3}$). The third-order time derivative of SMD, which indicates the rate at which the acceleration of the SMD changes.

\textit{Angular snap of SMD} ($\varsigma=d\xi/dt$, arcsnap day$^{-4}$). The fourth-order time derivative of SMD, which specifies the rate at which the jerk of SMD occurs.

\textit{Angular velocity of the EoT} ($\omega^*=d\delta^*/dt$, arcmin day$^{-1}$). The first-order time derivative of the EoT within a solar meridional analemma, which describes the rate of change for EoT.

\textit{Angular acceleration of the EoT} ($\alpha^*=d^2 \delta^*/dt^2$, arcsec day$^{-2}$). The second-order time derivative of the EoT within a solar meridional analemma, which quantifies the rate of change for the velocity of EoT.

\textit{Angular jerk of the EoT} ($\xi^*=d^3 \delta^*/dt^3$, arcjerk day$^{-3}$). The third-order time derivative of the EoT, which indicates the rate at which the acceleration of EoT changes.

\textit{Angular snap of the EoT} ($\varsigma^*=d^4 \delta^*/dt^4$, arcsnap day$^{-4}$). The fourth-order time derivative of the EoT, which accounts for the rate at which jerk occurs.

\textit{arcjerk}. A unit required to undertake this analysis. One arcjerk is equivalent to 1/60$^3$ arcdeg, 1/60$^2$ arcmin, 1/60 arcsec or 60 arcsnap.

\textit{arcsnap}. A unit required to undertake this analysis. 1 arcsnap is equivalent to 1/60$^4$ arcdeg, 1/60$^3$ arcmin, 1/60$^2$ arcsec and 1/60 arcjerk.

\textit{Crest}. A maximum of a function where the direction of the motion switches from upward to downward, (or from right to left). The slope of a function becomes zero at the crest.

\textit{Cycle}. Repetitive loop of a sinusoidal curve. The cycle is completed when final and initial conditions converge, connecting one cycle to the next.

\textit{Earth's rotational speed} ($\mathrm{ER}_\omega$, km h$^{-1}$). The linear velocity at which Earth's surface rotates relative to the solar ring, specifically for the Equator. 

\textit{Earth-Sun axis of gravity} (or helio-terric line of gravity) An axis that connects the Sun and Earth centers of mass and gravity, while meaning the shortest distance between their surfaces. Proposed here as the axis for the control of Earth's revolution and rotation, with the Sun as a sole driver.
 
\textit{Equation of Time, or EoT} ($\delta^*$). Also referred here as the Sun's horizontal path. The EoT is the time difference between the mean day and the solar day.

 \textit{Length of the solar day} (LSD) is a cycle that measures the period between two occurrences of NBI in the meridian of a given site of Earth's surface. It is closely linked to both $\delta^*$ and $\mathrm{ER}_\omega$. The three cycles are different approaches to the same phenomenon.

\textit{Light cone, or sunbeam cone}. A figurative conic section interrupted by two circular planes, the larger plane corresponds to the solar disk along its full two-dimensional shape on the sky, while the smaller plane is the lumbra.

\textit{Lumbra}. A circular area drawn when the light-cone lands on Earth's surface, such area holding natural beam irradiance (NBI) simultaneously. The center of the lumbra is the subsolar point.

\textit{Natural beam irradiance (NBI)}. The amount of solar irradiance (Nm$^{-2}$) delivered as a normal projection from the solar ring to the lumbra.

\textit{Net drive}. The net drive, or resultant drive, associates the sign of the velocity on which a motion points and the sign of its acceleration.

\textit{Period}. The time it takes for an entire cycle to be completed.

\textit{Phase of EoT} (or phase of the analemma). Each of 4 segments defined here for the EoT within a solar sundial noon analemma.

\textit{Pseudo-sinusoidal function}. A graph resembling a sinusoidal function, but whose amplitudes, periods and zero crossing points fail to occur on a neat rhythm.

\textit{Solar meridional analemma} ($\delta',\delta^*$). A function built by plotting SMD over the EoT. For any given site on Earth's surface, the solar meridional analemma displays the combined vertical-horizontal path of the mean-time Sun.

\textit{Solar sundial analemma}. When the sign of the EoT is switched, the analemma becomes reversed regarding the horizontal axis, conforming a sundial analemma.

\textit{Sun meridian declination, or SMD} ($\delta$). The Sun's vertical path or true declination. The SMD is the angle between the Sun and Earth's Equator.

\textit{Trough}. A minimum of a function, where the direction of the motion switches from downward to upward, or from left to right.

\textit{Zero crossing point (ZCP)}. When a function oscillates around zero, a ZCP is the point where the curve crosses the abscissa.

\section*{Data Availability}
Data is available as: 2025. https://doi.org/10.6084/m9.figshare.30389770
(last-time accessed on 20 Oct 2025).


\begin{thebibliography}{}
  \bibitem[\protect\citename{Athayde Jr, }2015]{Athayde2015}
   Athayde Jr, L. S. 2015, \textit{J. Aeronaut. Aerosp. Eng.}, 4, 1000145

  \bibitem[\protect\citename{Cambridge, }2025]{Cambridge2025}
   [dataset] Cambridge, Author 2025, FigShare
   
   \bibitem[\protect\citename
   {Camuffo {\it et al.}, }2021]{Camuffo2021}
   Camuffo, D., Della Valle, A. and  Becherini, F. 2021,\textit{Climatic Change, 165, 38}
  
  \bibitem[\protect\citename{Dickey, }1995]{Dickey1995}
   Dickey, J. O. 1995, in \textit{Global Earth Physics: A Handbook of Physical Constants}, ed. J. T. Ahrens (Washington D.C.: Am. Geophys. Union), 356
  
  \bibitem[\protect\citename{Georgieva, }2006]{Georgieva2006}
   Georgieva, K. 2006, in \textit{Space Science} (New York: Nova Sci. Inc.), 35
  
  \bibitem[\protect\citename{Gross, }2016]{Gross2016}
   Gross, R. S. 2016, \textit{Int. Assoc. Geod. Symp.}, Springer Verlag, 41, 13

  \bibitem[\protect\citename{Lynch, }2012]{Lynch2012}
   Lynch, P. 2012, \textit{The Equation of Time and the Analemma}, 
   \textit{Irish Math. Soc. Bulletin}, \textbf{69}, 47--56. 

   \bibitem[\protect\citename{Meeus, }1998]{Meeus1998}
   Meeus, J. 1998, \textit{Astronomical Algorithms}, 2nd ed. (Richmond, VA: Willmann–Bell)
    
  \bibitem[\protect\citename{Muller, }1995]{Muller1995}
   Muller, M. 1995, \textit{Acta Phys. Pol. A}, 88, 49
  
  \bibitem[\protect\citename{Nurick, }2011]{Nurick2011}
   Nurick, A. 2011, \textit{Sol. Energy}, 85, 295
  
  \bibitem[\protect\citename{Raisz, }1941]{Raisz1941}
   Raisz, E. 1941, \textit{J. Geogr.}, 40, 90
  
  
  \bibitem[\protect\citename{Rueda {\it et al.}, }2024]{Rueda2024}
   Rueda, J. A., Ramírez, S., Sánchez, M. A., and Guerrero, J. de D. 2024, \textit{Atmosphere (Basel)}, 15, 1003
  
  \bibitem[\protect\citename{Shaw, }2002]{Shaw2002}
   Shaw, S. G. 2002, \textit{Navig.}, 49, 1
  
  \bibitem[\protect\citename{Spencer, }1971]{Spencer1971}
   Spencer, J. W. 1971, \textit{Search}, 2, 172
   
  \bibitem[\protect\citename{Torge {\it et al.}, }2023]{Torge2023}
   Torge, W., Müller, J., and Pail, R. 2023, \textit{Geodesy}, 5th ed. (Berlin: De Gruyter Oldenbourg)
  
  \bibitem[\protect\citename{Williams, }n.d.]{WilliamsND}
   Williams, D. R. n.d., Earth Fact Sheet, NASA Goddard Space Flight Center, accessed Dec 20, 2024
\end{thebibliography}
\end{document}